\documentclass[10 pt]{article}
\usepackage{epsfig}
\begin{document}

\title{\bf The Quantum Interference Computer: an experimental proposal}

\author{A.Y. Shiekh\footnote{\rm shiekh@dinecollege.edu} \\
            {\it Din\'{e} College, Tsaile, Arizona, U.S.A.}}

\date{}

\maketitle

\abstract{An experiment is proposed to test the interference aspect of the Quantum Interference Computer approach}

\baselineskip .5 cm

\section{Introduction: Computing with the quantum}

The logical bit (binary digit) is the fundamental concept in classical digital
computing and can take on the state representing 0 or 1. In contrast, the world on a
small (atomic) scale obeys differing rules described by quantum theory, which has the
qubit that can be a linear superposition of these two states:
\begin{equation}
\left| \phi \right> = \alpha \left| 0 \right> + \beta  \left| 1 \right>
\end{equation}
a seemingly small change that has many profound consequences, where the amplitudes
$\alpha$ and $\beta$ are complex numbers, and are the analog part of quantum theory.
However, and in contrast, when we measure such a state we actually get the result 0 or
1 (the state has collapsed) with probabilities 
$|\alpha |^2$ for $\left| 0 \right>$ and $|\beta |^2$ for $\left| 1 \right>$, such is the
nature of the quantum world and this is the digital aspect of quantum theory
where conservation of the system (unitarity) demands that $|\alpha|^2 + |\beta|^2 = 1$.
Why the world is like this, nobody really knows, and it disturbed Einstein to such an
extent that he stated that `God does not play dice'; but without such a mechanism, we
would be denied freewill, so it is a good thing that the world is the way that it is.

In the large world such interactions are happening all the time, and that is why we
are not used to seeing the direct effect of these combinations. It takes a lot of care
to avoid a measurement happening until one is ready, and this is part of the
difficulty in building a quantum computing device.

So here we see the nature of the quantum way, where, although both bit types are
involved, only one is seen upon measurement. There are features of an analog 
system (the continuous numbers $\alpha$ and $\beta$), 
while the act of measurement carries discrete, or digital, aspects.

We have avoided delving on the more subtle and strange aspects of quantum theory at
this juncture, and if necessary one can adopt a pragmatic Engineering approach.

\subsection{The parallel nature of quantum theory}

Because the quantum state carries both digits at once, unlike the classical, there is
the prospect of performing many calculations in parallel. This is seen even more
clearly for a 2 qubit quantum system, whose state would look like:
\begin{equation}
\left| \psi \right> = 
\alpha_{00} \left| 0 \right> \left| 0 \right> + 
\alpha_{01} \left| 0 \right> \left| 1 \right> +
\alpha_{10} \left| 1 \right> \left| 0 \right> +
\alpha_{11} \left| 1 \right> \left| 1 \right>
\end{equation}
while in general an $n$ qbit system has $2^n$ components. This exponential growth in
size is at the potential core of the power implicit to quantum computing, and is of
such an enormous advantage that a system with just 300 bits would have more states
than there are atoms in the visible Universe (about $10^{80}$). This leads one to
pondering how or where all these calculations are performed and held, and such questions
remind us why Physics once went under the name of Natural Philosophy.

The above state is often written more compactly as:
\begin{equation}
\alpha_{00} \left| 0 0 \right> + 
\alpha_{01} \left| 0 1 \right> +
\alpha_{10} \left| 1 0 \right> +
\alpha_{11} \left| 1 1 \right>
\end{equation}
and if one were then to apply a function ($f$) to this one state, Nature's quantum
engine would effectively apply it to all components, yielding:
\begin{equation}
\alpha_{00} f(\left| 0 0 \right>) + ... +
\alpha_{11} f(\left| 1 1 \right>)
\end{equation}

The ability to do so very much computing in one application is the good part; how this
is actually achieved by Nature is not known.

\subsection{The restrictions of measurement}

The problem (or bad part) arises upon the act of measurement, when, as mentioned
above, one only sees one of the parts with due probability. As a result no advantage
has been taken of the fact that the quantum world has all that computational power,
and this is exactly why quantum computers seem to be so hard to program.

Rather than detour at this point into a discussion of the various restricted approaches
to date known to overcome this obstacle, we consider an alternative proposal that might
show promise of a generic way around this dilemma.

\section{A review of the Quantum Interference approach}

Interference has been proposed as an amplifying mechanism for quantum computation \cite{Shiekh, Shiekh2, Long}. How it is supposed to work is illustrated by the following.

Start with the following three qubit Hadamard state for illustration (leaving
out normalizations for clarity)
\begin{eqnarray}
\left| \psi  \right> &=&
(\left| 0 \right> + \left| 1 \right>)
(\left| 0 \right> + \left| 1 \right>)
(\left| 0 \right> + \left| 1 \right>) \\
&=&
 \left| 0 0 0 \right> + 
 \left| 0 0 1 \right> +
 \left| 0 1 0 \right> +
 \left| 0 1 1 \right> + \ldots +
 \left| 1 1 1 \right>
\end{eqnarray}
and like Grover's algorithm, apply the decision function to mark the invalid
solutions by inverting their phase. For the sake of argument let us suppose that
the solutions 001 and 011 satisfy the function, which yields the state:
\begin{equation}
 - \left| 0 0 0 \right> + \overbrace{\bf \left| 0 0 1 \right>}^{\it Solution}
 - \left| 0 1 0 \right> + \overbrace{\bf \left| 0 1 1 \right>}^{\it Solution} - \ldots
 - \left| 1 1 1 \right>
\end{equation}
which has got us nowhere at all, {\it unless} one were to bring in the mechanism of
Young's double slit or the beam splitter interferometer, with the marking function
being applied to one of the two arms alone. Then interference of the arms would
yield:
\begin{eqnarray}
\begin{array}{cc}
 &- \left| 0 0 0 \right> + {\bf \left| 0 0 1 \right>}
   - \left| 0 1 0 \right> + {\bf \left| 0 1 1 \right>} - \dots
   - \left| 1 1 1 \right>
\\
 &+ \left| 0 0 0 \right> + {\bf \left| 0 0 1 \right>}
   + \left| 0 1 0 \right> + {\bf \left| 0 1 1 \right>} + \ldots
   + \left| 1 1 1 \right>
\end{array}
\end{eqnarray} 
to expose the desired solutions
\begin{equation}
{\bf \left| 0 0 1 \right>} + {\bf \left| 0 1 1 \right>}
\end{equation}
one of which will consolidate upon measurement, and can then be confirmed on a classical
computer, if so desired. The two arms are brought into overlap and not sent through a
final beam splitter as is typical of an interferometer.

To locate the remaining solution, one can start over, and exclude the known solution
by also inverting its phase in one of the two interference arms. Eventually all
solutions will be located and removed, so the final run will expose either a non-valid
solution or a previously found solution from the remnants of the wave-function.

Concerns over lost unitarity can be allayed by noting that a quantum computer
typically starts by transforming a sharp (ground) state into a superposition, and that
this is a unitary change. All that is happening here is the inverse, and so the
process is also unitary.

In practice, due to imperfect cancellation, this process may need to be repeated a few
times before the act of measurement.

\section{The interference question}
The most questionable aspect of the quantum interference approach is the use of interference to amplify out the correct solution, and this could be experimentally tested for the simplest case of just two candidate solutions, where only one is to survive.

Some people will have concerns over the use of destructive quantum interference where it
might initially seem possible to `destroy' the wave function by trying to arrange for total
destructive interference between the two arms
\begin{equation}
\left| \phi \right> - \left| \phi \right>
\end{equation}
but one must not forget that this is a physical process that can actually be performed in
reality, although never with perfection, and as a result any remnant error will be
reunitarized by Nature, since all quantum mechanical processes are norm-preserving. It is
this very mechanism of magnification that is extracting the solutions in the quantum
interference proposal.

There seems no reason to believe that quantum theory itself might need modifying to cover
this situation.

\subsection{The experimental proposal}

Start with vertically polarized light
\begin{equation}
\left| 1 \right>
\end{equation}
and split this into two arms using a beam splitter, or even just a double slit; place a polarizer in each arm\footnote{Optical activity can be used to achieve this rotation with minimal losses}, each diagonally aligned relative to the original beam, but perpendicular to each other. 

\begin{center}
\includegraphics[scale=.5]{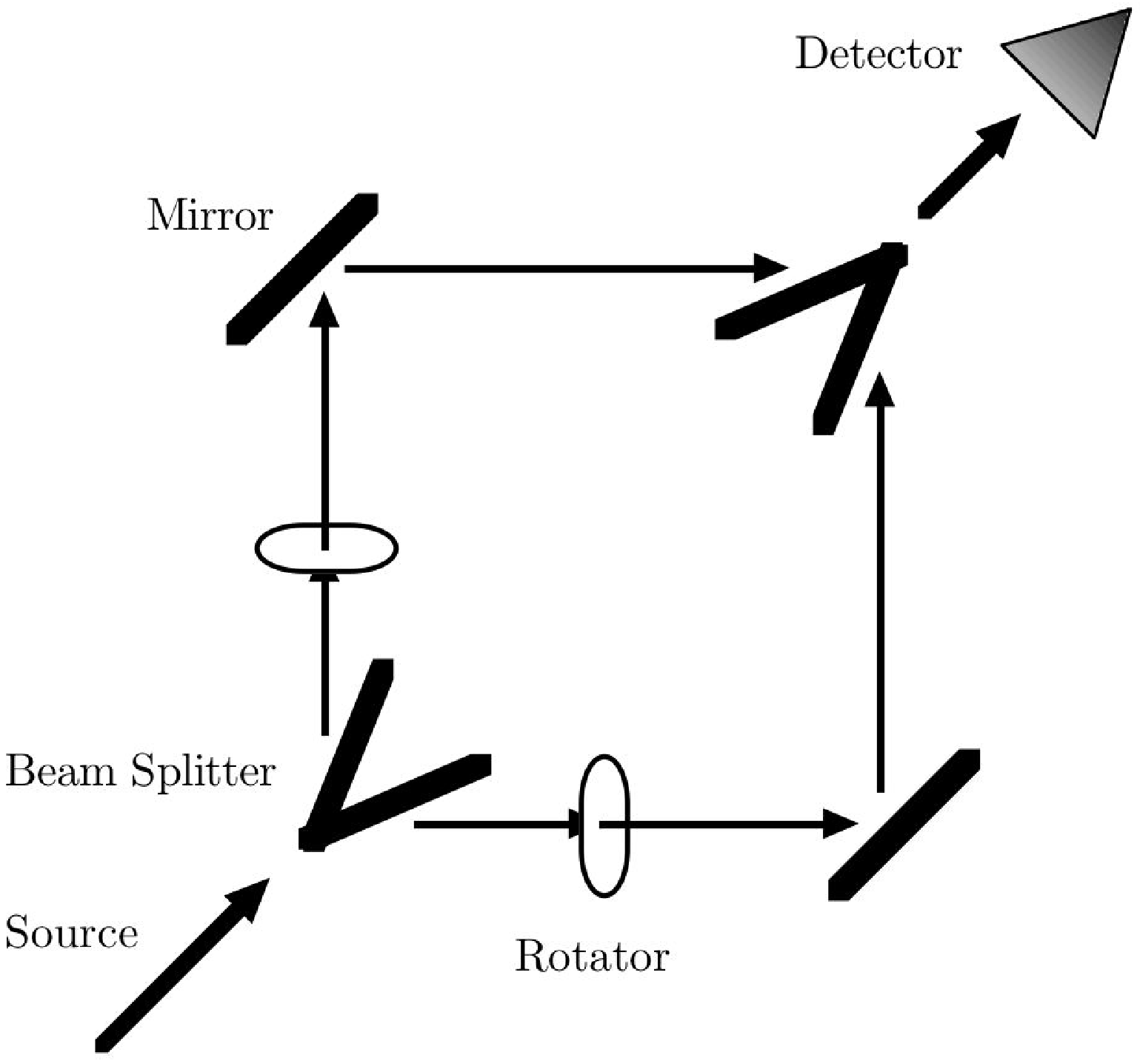}
\end{center}
\begin{center}
Test Interferometer
\end{center}

For cases where the wave function makes it through (and does not collapse), the state of each arm would then be
\begin{equation}
\left| 0 \right> + \left| 1 \right>
\end{equation}
and
\begin{equation}
\left| 0 \right> - \left| 1 \right>
\end{equation}
in this way simulating the marking of state $\left| 1 \right>$ as invalid. The two arms should then be brought into physical overlap, where they will undergo interference to yield a predicted horizontal polarization.
\begin{equation}
 \left| 0 \right>
\end{equation}

This experiment should be a way of testing the amplification ability of interference.

\section{Conclusion}
These experiment should test the interference and re-normalization effects of quantum theory.


\begin{thebibliography}{99}
\bibitem{Shiekh} A. Y. Shiekh, Int. Jour. Theo. Phys., 45, 2006, 1653 [arXiv:cs.CC/0507003]
\bibitem{Shiekh2} A. Y. Shiekh [ariXiv:quant-ph/0611052]
\bibitem{Long} Gui Lu Long, Commun. Theor. Phys. 45(5), 2006, 825 [ariXiv:quant-ph/0512120]
\end{thebibliography}
\end{document}